\documentclass[prd,twocolumn,showkeys,showpacs,amsmath,amssymb,superscriptaddress,preprintnumbers,nofootinbib]{revtex4}
\usepackage{amssymb}
\usepackage{amsmath}
\usepackage{graphicx}

\normalbaselines
\begin{document}

\title{Electrodynamic phenomena induced by a dark fluid: \\ Analogs of pyromagnetic, piezoelectric, and striction effects}

\author{Alexander B. Balakin}
\email{Alexander.Balakin@kpfu.ru}
\author{Nadejda N. Dolbilova}
\email{nadejda.dolbilova@gmail.com}
\affiliation{Department of General Relativity and
Gravitation, Institute of Physics,Kazan Federal University, Kremlevskaya street 18, 420008, Kazan,
Russia}

\begin{abstract}
We establish a new model of coupling between a cosmic dark fluid
and electrodynamic systems, based on an analogy with effects of
electric and magnetic striction, piezo-electricity and
piezo-magnetism,  pyro-electricity and pyro-magnetism, which
appear in classical electrodynamics of continuous media. Extended
master equations for electromagnetic and gravitational fields are
derived using Lagrange formalism. A cosmological application of
the model is considered, and it is shown that a striction-type
interaction between the dark energy (the main constituent of the
dark fluid) and electrodynamic system provides the universe
history to include the so-called unlighted epochs, during which
electromagnetic waves can not propagate and thus can not scan the
universe interior.
\end{abstract}



\maketitle

\section{Introduction}

Dark fluid composed of a dark energy  and a dark matter is
considered nowadays as a key constitutive element of modern
cosmological models (see, e.g.,
\cite{DE1,DE2,DE3,DM1,DM2,DM3,DMDE,DEcosmo,DE2011,DEmodified,DF1,DF2,DF3,DF4,DF99}).
Both the dark energy and dark matter are assumed to consist of
electrically neutral particles and thus the dark fluid does not
interact with an electromagnetic field {\it directly}. That is
why, we would like, first of all, to explain the terminological
context, which allows us to speak about electrodynamic phenomena
induced by the dark fluid. Let us imagine a hierarchical
cosmological system, in which the dark fluid (energetically
dominating substrate with a modern contribution about $95\%$) is
considered to be the guiding element, and an electrodynamic
subsystem (as a part of baryonic matter with its modern
contribution about $5\%$) to be the subordinate element. In this
context the electrodynamic subsystem plays the role of a {\it
marker}, which signalizes about the variations in the state of
dark fluid, the energetic reservoir, into which this marker is
immersed. We are interested to answer the question: what
mechanisms might be responsible for a (possible) transmission of
information about the dark fluid state to the electrodynamic
system. Since electrodynamic systems form the basis for the most
important channel of information about the universe structure, one
could try to reconstruct features of the dark fluid evolution by
tracking down specific fine details of the spectrum of observed
electromagnetic waves, of their phase and group velocities.

The most known {\it marker-effect} of such type is the
polarization rotation of electromagnetic waves travelling through
the axionic dark matter \cite{PR1,PR2,PR3,PR5}. Let us remind a
few details of this phenomenon. From the physical point of view,
the corresponding mechanism is connected with magneto-electric
cross-effect \cite{Dell,HehlObukhov}, which is generated in the
medium by the pseudoscalar field associated with dark matter
axions. From the mathematical point of view, this mechanism is
described by inserting a special term into the Lagrangian,
$\frac14 \phi F^*_{mn}F^{mn}$, which is linear in the pseudoscalar
(axion) field $\phi$ and is proportional to the pseudo-invariant
of the electromagnetic field quadratic in the Maxwell tensor
\cite{ax1}. The model of this axion-photon coupling was extended
for the non-stationary state of the dark matter (see, e.g.,
\cite{BBT1,BBT2}), and for the states, for which nonminimal
effects linear in the space-time curvature are significant (see,
e.g., \cite{BNi}).

Availability of the example of the coupling of photons with the
dark matter axions encourages us to search for marker-effects
related to the interaction of electrodynamic system with the dark
energy, the main constituent of the dark fluid. We assume that
electrodynamic systems can be influenced by the pressure of the
dark energy in analogy with mechanical stresses, which are known
to control the response in electric and magnetic materials in
industry and technique. To be more precise, we can search for dark
fluid analogies with the following classical effects. First, we
mean the analogy with the classical piezo-electric effect (the
appearance of an electric polarization in the medium influenced by
mechanical stress and vice-versa), and with the classical
piezo-magnetic effect (appearance of a magnetization under
stress)(see, e.g., \cite{Nye,SSh,LL,Mauginbook}). Second, we would
like to consider an analogy with the inverse electrostriction
effect (a combination of external pressure and electric field
generates the electric polarization in the medium), and an analogy
with the inverse magnetostriction effect (a combination of
external pressure and magnetic field generates the magnetization
in the medium), as well as, an analogy with the magneto-electric
cross-effect displayed by the external stress. Based on results of
classical electrodynamics of continuous media, we can expect that
piezo-effects will be visualized, when the dark fluid is
anisotropic (e.g., in the early universe). The striction effects
due to their symmetries are expected to be available in the
isotropic universe also. In addition to piezo- and striction-
effects induced by the dark fluid pressure we can expect the
appearance of marker-signals similar to pyro-electric and
pyro-magnetic responses of the medium, in which the temperature
changes with time \cite{Nye,SSh,LL} (pyro-effect also is hidden,
when the dark fluid is isotropic). One of the important
characteristic of the dark fluid is its macroscopic velocity
four-vector and covariant derivative of this four-vector. When we
focus on the influence of the dark fluid non-uniform motion on the
properties of electrodynamic system, we, in fact, search for
analogs of dynamo-optical phenomena \cite{LL}; we hope to consider
these phenomena in detail in the next paper. In principle, we
could consider an analogy with the so-called thermo-electric and
thermo-magnetic effects, induced by heat-fluxes in the medium, but
this sector of physical modeling is out of scope of this paper.
Also, in this paper we do not consider magneto-electric
cross-effects induced by the combination of the dark energy
pressure and of the axionic dark matter. Effects of this type are
worthy of special consideration.

We have to emphasize that mathematical theory of pyro-, piezo- and
striction- effects is developed in detail  for classical
electrodynamics of continuous media, and below we consider a
general relativistic extension of that theory for the case of dark
fluid action on the electrodynamic system. In this sense, we take
the mathematical scheme of the description of such interactions,
which is well-tested, has clear interpretation and is based on the
Lagrange formalism, and then we construct its general relativistic
analog, using this scheme in the context of dark fluid
electrodynamics.

This paper is organized as follows. In Section II we remind the
terminology and introduce the Lagrangian and master equations for
the model of electromagnetically inactive dark fluid. Section III
contains detailed description of the extended model: in Section
III.A we extend the Lagrangian by the terms, which describe
interactions of the pyro-, piezo- and striction- types between
dark energy and electrodynamic system; in Section III.B we derive
extended electrodynamic equations and discuss the structure of
tensor coefficients describing pyro- (III.B.1), piezo- (III.B.2)
and striction-(III.B.3) coefficients associated with the coupling
to dark energy; in Section III.C we obtain the extended gravity
field equations in general form. In Section III.D we write the
equation for a axion field attributed to the dark matter. In
Section IV we reduce the derived master equations to the case,
when the medium is spatially isotropic: Section IV.A contains
details of reduced electrodynamic equations; in Section IV.B we
collect details of modified gravity field equations; in Section
IV.C we consider an example of extended model with hidden magnetic
and/or electric anisotropy. In Section V we consider a
cosmological application of the established model to the problem
of description of the so-called unlighted epochs in the universe
history, and their relations to the striction-type interactions of
electrodynamic systems with the cosmic dark energy. In Section VI
we summarize the results. Appendix includes working formulas for
the extended variation procedure.

\section{Electromagnetically inactive dark fluid}

In order to remind the standard elements of the theory and to introduce new details,
let us start with the model, in the framework of which the electromagnetic field interacts with the standard matter only, and
the dark fluid is coupled with the electrodynamic system by the gravitational field only.

\subsection{The Lagrangian}

Let us remind  that the standard Einstein-Maxwell model is described by the action functional
\begin{equation}
S_{0} {=} \int d^4 x \sqrt{{{-}}g} \left[\frac{R}{2\kappa} {+}
L_{({\rm DF})} {+} \frac{1}{4} C_{(0)}^{ikmn} F_{ik}F_{mn} {+}
L_{({\rm m})} \right], \label{0actmin}
\end{equation}
which is quadratic in the Maxwell tensor $F_{ik}$. Here $g$ is the
determinant of the metric tensor $g_{ik}$, the term $R$ is the
Ricci scalar, $\kappa {=} \frac{8 \pi G}{c^4}$ is the Einstein
constant, $L_{({\rm DF})}$ is the Lagrangian of the dark fluid,
$L_{({\rm m})}$ is the Lagrangian of a standard matter. The tensor
$C^{ikmn}_{(0)}$ describes standard linear electromagnetic
response of the medium formed by the standard matter; in vacuum
the corresponding term takes the form $\frac{1}{4} F^{mn}F_{mn}$.
As usual, we assume that $L_{({\rm m})}$ does not include the
Maxwell tensor $F_{mn}$, nevertheless, it can depend on the
potential four-vector $A_i$, if the medium is conductive.

\subsection{Standard electrodynamic equations}

The Maxwell tensor is represented in terms of a four-vector
potential $A_i$ as
\begin{equation}
F_{ik} = \nabla_i A_{k} - \nabla_k A_{i} \,, \label{maxtensor}
\end{equation}
and thus satisfies the condition
\begin{equation}
\nabla_{k} F^{*ik} =0 \,, \label{Emaxstar}
\end{equation}
where $F^{*ik} \equiv \frac{1}{2} \epsilon^{ikpq}F_{pq}$ is the
tensor dual to $F_{pq}$, the term $\epsilon^{ikpq} \equiv
\frac{1}{\sqrt{-g}} E^{ikpq}$ is the Levi-Civita tensor,
$E^{ikpq}$ is the absolutely skew-symmetric Levi-Civita symbol
with $E^{0123}=1$. The variation of the action functional
(\ref{0actmin}) with respect to the four-vector potential $A_i$
gives the following  electrodynamic equations
\begin{equation}
\nabla_{k} H^{ik} = - \frac{4\pi}{c} I^i \,, \quad H^{ik} =
C_{(0)}^{ikmn}F_{mn} \,. \label{induc1}
\end{equation}
Here $H^{ik}$ is the excitation tensor \cite{HehlObukhov}, and
the four-vector $I^i$ defined as
\begin{equation}
I^i = \frac{1}{4\pi} \frac{\delta L_{({\rm m})}}{\delta A_i} \,,
\quad \nabla_i I^i = 0 \,, \label{E8}
\end{equation}
describes the electric current.
In this paper we consider the medium to be non-conducting, i.e., $I^i {=}0$.

\subsection{Gravity field equations}

Variation of the action functional (\ref{0actmin}) with respect to
metric gives the equations of the gravitational field, which can
be written in the following form
\begin{equation}
R_{ik}-\frac{1}{2}Rg_{ik} = \kappa \left[ T^{(0)}_{ik} + T^{({\rm
DE})}_{ik} + T^{({\rm DM})}_{ik} + T^{({\rm m})}_{ik}\right] \,. \label{EineqMIN}
\end{equation}
Here $T^{(0)}_{ik}$ is the effective symmetric traceless
stress-energy tensor of electromagnetic field in a continuous
medium
\begin{equation}
T^{(0)}_{ik} \equiv  \left[\frac{1}{4}g_{ik} F_{mn} {-} \frac12
\left(g_{im}F_{kn}{+}g_{km}F_{in} \right)\right]
C^{mnpq}_{(0)}F_{pq}\,, \label{TEM}
\end{equation}
(see, e.g., \cite{2007A,2007B} for details). The Lagrangian of the electromagnetically inactive dark fluid is presented as a sum
$L_{({\rm DF})} \to L_{({\rm DE})} {+}L_{({\rm DM})}$, and the corresponding stress-energy tensors enter the right-hand side of
equations (\ref{EineqMIN}) also as the sum.
The stress-energy tensor of
the dark energy is defined as
\begin{equation}
T^{({\rm DE})}_{ik} = - \frac{2}{\sqrt{-g}} \frac{\delta
\left[\sqrt{-g} \ L_{({\rm DE})}\right]}{\delta g^{ik}} \,.
\label{TAX}
\end{equation}
Let us define the unit four-vector $U^i$  of the macroscopic
velocity as an eigen-vector of the stress-energy tensor of the
dark energy, i.e., let us assume that $T^{({\rm DE})}_{ls} U^s {=}
W U_l$, $U^iU_i{=}1$ (the so-called Landau-Lifshitz definition).
Then this tensor can be algebraically represented in the following
form
\begin{equation}
T^{({\rm DE})}_{ik} \equiv W U_i U_k + {\cal P}_{ik}\,,
\label{fluid}
\end{equation}
where the eigen-value $W$ is interpreted as the energy density of
the dark energy, and ${\cal P}_{ik}$ is its pressure tensor. These
quantities can be written as follows:
\begin{equation}
W = U^l T^{({\rm DE})}_{ls} U^s \,,  \quad {\cal
P}_{ik} = \Delta^l_i T^{({\rm DE})}_{ls} \Delta_k^s \,,
\label{1fluid}
\end{equation}
where $\Delta^l_i \equiv \delta^l_i {-}U^lU_i$ is the projector.
The tensors $T^{({\rm DM})}_{ik}$ and $T^{({\rm m})}_{ik}$, which describe the contributions of the dark matter
and standard matter, respectively, can be obtained by the formulas similar to (\ref{TAX}), however, their algebraic
decompositions are more sophisticated, since the macroscopic
velocity four-vector $U^i$ is already fixed as an eigen-vector of
the dark energy stress-energy tensor. For instance, $T^{({\rm m})}_{ik}$ is of the form
\begin{equation}
T^{({\rm m})}_{ik} \equiv W^{({\rm m})} U_i U_k + U_i I_k^{({\rm
m})} + U_k I_i^{({\rm m})} + P^{({\rm m})}_{ik}\,, \label{Tmatter}
\end{equation}
and includes the heat-flux four-vector $I_k^{({\rm m})} \equiv
\Delta^l_k T^{({\rm m})}_{ls} U^s$ of the matter in addition to
the matter energy-density $W^{({\rm m})}$ and the matter pressure
tensor $P^{({\rm m})}_{ik}$. When we describe the dark matter, one should
change the symbol $({\rm m})$ by $({\rm DM})$ in the formula (\ref{Tmatter}).

\section{Electromagnetically active dark fluid}

\subsection{Extended Lagrangian}

Let us extend the action functional to include the terms
describing the interaction between the dark fluid and electrodynamic system.
We assume the extension to have the following  form
$$
S {=} \int d^4 x \sqrt{{-}g} \left\{\frac{R}{2\kappa} {+} L_{({\rm
DE})} {+} \frac{1}{4} C_{(0)}^{ikmn} F_{ik}F_{mn}  {+} L_{({\rm m})}
{+} \right.
$$
$$
\left.
{+} \frac12 \Psi^2_0 \left[- \nabla_k \phi \nabla^k \phi + {\cal V}(\phi^2) \right] {+} \frac14 \phi F^*_{mn} F^{mn} {+}
\right.
$$
$$
\left.
{+} \frac12 \left( \pi^{ik} {+} \frac12 \lambda^{ikmn} F_{mn} \right)F_{ik}  DW {+}
\right.
$$
\begin{equation}
\left.
{+} \frac12 \left({\cal D}^{ikpq} {+} \frac{1}{2} \
Q^{ikmnpq} F_{mn} \right) F_{ik} \ {\cal P}_{pq} \right\}\,, \label{actmin}
\end{equation}
which again is up to second order in the Maxwell tensor $F_{ik}$,
but in addition to the second order terms also the terms linear in
$F_{mn}$ appeared. Here and below we use the symbol $D$ for the
convective derivative $D \equiv U^i \nabla_i$. We specified the
Lagrangian of the dark matter $L_{({\rm DM})}$ as the one for a
pseudoscalar (axion) field $\phi$; in this context the quantity
$\frac{1}{\Psi_0}$ is a coupling constant of the axion-photon
interaction, and ${\cal V}(\phi^2)$ is the potential of the
pseudoscalar field. In this terms, the cross-invariant $\frac14
\phi F^*_{mn} F^{mn}$ describes the coupling between the
electromagnetic and pseudoscalar fields, i.e., the interaction
between axionic dark matter and electromagnetic field \cite{ax1}.
In other words, we deal here with the example of
electromagnetically active dark fluid, and this type of activity
is connected with the dark matter part of the dark fluid.

Now we consider the dark fluid activity related to the interaction
of the electromagnetic field with dark energy constituent of the
dark fluid. Based on the analogy with electrodynamics of
continuous media we can consider the first term linear in the
Maxwell tensor, $\frac12 \pi^{ik}F_{ik} DW$, as describing the
analog of the pyro effects (pyro-electric and/or pyro-magnetic).
Of course, in classical electrodynamics of continuous media one
deals with the convective derivative of the temperature $DT$, when
one speaks about pyro - effects, nevertheless, assuming that
$DW{=}\frac{dW}{dT} DT$ we keep this terminology for the dark
fluid also and indicate the tensor $\pi^{ik}$ as the tensor of
pyro-coefficients. The second term linear in the Maxwell tensor,
$\frac12{\cal D}^{ikpq}F_{ik} {\cal P}_{pq}$, includes the
pressure tensor of the dark energy ${\cal P}_{pq}$ and thus
describes analogs of piezo-effects (piezo-electric and/or
piezo-magnetic). The corresponding piezo-coefficients are encoded
in the tensor ${\cal D}^{ikpq}$. The term quadratic in $F_{mn}$
and linear in $DW$ describes the part of the linear
electromagnetic response, which depends on the rate of evolution
of the energy density of the dark fluid. The last term in
(\ref{actmin}) is quadratic in the Maxwell tensor and linear in
the pressure tensor, thus describing  the response associated with
electro- and magneto-striction, induced by the dark energy; the
tensor $Q^{ikmnpq}$ introduces coefficients of electro- and
magneto-striction.

\subsection{Extended electrodynamic equations}

The variation of the action functional
(\ref{actmin}) with respect to the four-vector potential $A_i$
gives the following  electrodynamic equations
\begin{equation}
\nabla_{k} H^{ik} = - \frac{4 \pi}{c} I^i \,, \quad H^{ik} = {\cal
H}^{ik} {+} \phi F^{*ik} {+} {\cal C}^{ikmn}F_{mn} \,. \label{1induc1}
\end{equation}
Here $H^{ik}$ is the extended excitation tensor.
The term
\begin{equation}
{\cal H}^{ik} \equiv \pi^{ik} DW + {\cal D}^{ikpq}{\cal P}_{pq}
\label{2induc11}
\end{equation}
does not contain $F_{ik}$ and thus it can be indicated as the spontaneous
polarization-magnetization tensor. The term $\phi F^{*ik}$ is typical for the axion electrodynamics
(see, e.g., \cite{ax1}); the four-divergence of this term can be expressed as $F^{*ik}\nabla_k \phi$ due to (\ref{Emaxstar}).
Finally, the term
\begin{equation}
{\cal C}^{ikmn} \equiv
C^{ikmn}_{(0)} + \lambda^{ikmn} DW + Q^{ikmnpq} {\cal P}_{pq}
 \label{C1}
\end{equation}
describes the total linear response of the electrodynamic system including pyro-type and striction-type effects induced by the dark energy.

\subsubsection{Pyro- coefficients \\ associated with the coupling to dark energy}

The skew-symmetric tensor $\pi^{ik}$ describing the  pyro- effects
can be represented as
\begin{equation}
\pi^{ik}= \pi^i U^k - \pi^k U^i - \epsilon^{ik}_{\ \ mn} \mu^{m}
U^n \,,
 \label{pi03}
\end{equation}
thus visualizing the pyro-electric  $\pi^i$ and pyro-magnetic
$\mu^m$ coefficients, which are orthogonal to the velocity
four-vector ($\pi^iU_i {=} 0 {=} \mu^m U_m$). In general case the
dark energy can be characterized by three pyro-electric and three
pyro-magnetic coefficients. When we consider the dark energy as a
spatially isotropic medium, all six pyro coefficients vanish. For
the dark energy with an axial symmetry (e.g., in rotationally
symmetric Bianchi-I model) there are two non-vanishing
piezo-constants: one piezo-electric and one piezo-magnetic (see,
e.g., \cite{Nye,SSh} for details).

\subsubsection{Piezo-coefficients \\ attributed to the coupling to dark energy}

The tensor ${\cal D}^{ikpq}$ possesses the following symmetry of indices
\begin{equation}
{\cal D}^{ikpq} = - {\cal D}^{kipq} = {\cal D}^{ikqp} \,.
 \label{d1}
\end{equation}
Since the symmetric pressure tensor ${\cal P}_{pq}$ is considered to be orthogonal to the velocity
four-vector $U^i$, one can conclude that
\begin{equation}
{\cal D}^{ikpq} U_p = 0 = {\cal D}^{ikpq} U_q \,.
 \label{d2}
\end{equation}
This means that there are $6 \times 6 = 36$ independent coupling constants in the tensor of piezo-coefficients.
This tensor can be decomposed with respect to irreducible parts as follows:
\begin{equation}
{\cal D}^{ikpq} = d^{i(pq)} U^k - d^{k(pq)} U^i - \epsilon^{ik}_{\ \ ls}U^s h^{l(pq)}  \,.
 \label{d3}
\end{equation}
Here the piezo-electric coefficients $d^{i(pq)}$ and piezo-magnetic coefficients $h^{l(pq)}$ are defined by
\begin{equation}
d^{i(pq)} \equiv {\cal D}^{ikpq} U_k \,, \quad h^{l(pq)} \equiv \frac12 \epsilon^{ls}_{\ \ ik} {\cal D}^{ikpq} U_s \,,
 \label{d4}
\end{equation}
they are symmetric with respect to the indices $p,q$, and  are pure space-like, i.e., they satisfy the equalities
\begin{equation}
d^{i(pq)} U_i {=} 0 {=}  d^{i(pq)} U_p  \,, \quad h^{l(pq)} U_l {=}  0 {=} h^{l(pq)} U_p \,.
 \label{d5}
\end{equation}
In other words, in general case, the dark energy influence can be
characterized by 18 piezo-electric coefficients $d^{i(pq)}$ and/or
by 18 piezo-magnetic coefficients $h^{l(pq)}$. When the dark
energy is spatially isotropic, all these coefficients are equal to
zero. For the dark energy with axial symmetry there are four
non-vanishing piezo-electric and four non-vanishing piezo-magnetic
coefficients (see, e.g., \cite{Nye,SSh} for details).

\subsubsection{Permittivity tensors \\ associated with the coupling to dark energy}

Using the medium velocity four-vector $U^i$ one can decompose $ {\cal C}^{ikmn}$
uniquely as
$$
{\cal C}^{ikmn} =  \left( \varepsilon^{i[m}U^{n]} U^k  {-}
\varepsilon^{k[m} U^{n]} U^i \right) {-}
$$
\begin{equation}
{-}\frac12 \eta^{ikl}(\mu^{-1})_{ls}  \eta^{mns} {+}
\eta^{ikl}U^{[m} \nu_{l}^{\ n]} {+} \eta^{lmn}U^{[i}
\nu_{l}^{\ k]} \,.  \label{444}
\end{equation}
Here $\varepsilon^{im}$ is the tensor of total dielectric permeability,
$(\mu^{-1})_{pq}$ is the tensor of total magnetic impermeability, and
$\nu_{p \ \cdot}^{\ m}$ is the total tensor of magneto-electric
cross-effect induced by dark energy.  Keeping in mind the decomposition (\ref{C1}) one can
divide these quantities into three parts
$$
\varepsilon^{im} = 2 {\cal C}^{ikmn} U_k U_n =
\varepsilon_{(0)}^{im} {+} \sigma^{im} DW {+} \alpha^{im(pq)}
{\cal P}_{pq}  \,,
$$
$$
(\mu^{-1})^{ab}  = - \frac{1}{2} \eta^a_{\ ik} {\cal C}^{ikmn}
\eta_{mn}^{\ \ \ b} =
$$
$$
= (\mu^{-1})_{(0)}^{ab} + \rho^{ab} DW + \beta^{ab(pq)} {\cal
P}_{pq}\,,
$$
\begin{equation}
 \nu^{am} {=} \eta^a_{\ ik} {\cal C}^{ikmn} U_n {=} \nu^{am}_{(0)}{+} \omega^{am}DW {+} \gamma^{am(pq)} {\cal P}_{pq}\,. \label{varco}
\end{equation}
To complete the description of the tensor of cross-effects, we can write the sum of  $\nu^{am}$
related to the dark energy contribution (\ref{varco}) and
$\nu^{am}_{({\rm DM})} {=} \phi \Delta^{am}$ related to the contribution of the axionic dark matter.
In the formulas written above we introduced the corresponding spatial tensors as follows:
$$
\varepsilon_{(0)}^{im} = 2 C_{(0)}^{ikmn} U_k U_n \,, \quad \sigma^{im} = 2\lambda^{ikmn} U_k U_n \,,
$$
$$
\alpha^{im(pq)} = 2Q^{ikmnpq} U_k U_n \,,
$$
$$
(\mu^{-1})^{ls}_{(0)} = - \frac12 \eta^l_{\ ik} C_{(0)}^{ikmn}\eta^s_{\ mn}\,,
$$
$$
\rho^{ls} = - \frac12 \eta^l_{\ ik} \lambda^{ikmn}\eta^s_{\ mn} \,,
$$
$$
\beta^{ls(pq)} = {-} \frac12 \eta^l_{\ ik} Q^{ikmnpq}\eta^s_{\ mn} \,, \quad \nu^{am}_{(0)} {=} \eta^a_{\ ik} C_{(0)}^{ikmn}U_n \,,
$$
\begin{equation}
\omega^{am} {=} \eta^a_{\ ik} \lambda^{ikmn}U_n  \,, \quad \gamma^{lm(pq)} {=} \eta^l_{\ ik} Q^{ikmnpq}U_n \,. \label{0nu}
\end{equation}
As usual, the tensors $\eta_{mnl}$ and $\eta^{ikl}$ are the
skew-symmetric and orthogonal to $U^i$; they are defined as
\begin{equation}
\eta_{mnl} \equiv \epsilon_{mnls} U^s \,, \quad \eta^{ikl} \equiv
\epsilon^{ikls} U_s \,. \label{47}
\end{equation}
These tensors are connected by the useful identity
\begin{equation}
- \eta^{ikp} \eta_{mnp} = \delta^{ikl}_{mns} U_l U^s = \Delta^i_m
\Delta^k_n - \Delta^i_n \Delta^k_m \,. \label{usefulidentity}
\end{equation}
Upon contraction, equation (\ref{usefulidentity}) yields another
useful identity
\begin{equation}
\frac{1}{2} \eta^{ikl}  \eta_{klm} = - \delta^{il}_{ms} U_l U^s =
- \Delta^i_m \,. \label{50}
\end{equation}
The quantities $\delta^{ikl}_{mns}$ and $\delta^{il}_{ms}$ are the
generalized Kronecker deltas. Clearly, the two-indices tensors
$\varepsilon^{im}_{(0)}$, $(\mu^{-1})^{(0)}_{ab}$, $\sigma^{im}$,
$\rho^{ab}$ are symmetric and orthogonal to $U_k$; each of them
possesses six independent components. The spatial (pseudo)tensors
$\nu^{am}_{(0)}$ and $\omega^{am}$ are non-symmetric and thus each
of them contains nine independent components.

The spatial tensor $\alpha^{im(pq)}$ possesses the symmetry
\begin{equation}
\alpha^{im(pq)} = \alpha^{mi(pq)} = \alpha^{im(qp)} \,,
\label{alpha}
\end{equation}
and, generally, it has $6 \times 6=36$ independent components.
When the medium is spatially isotropic there are only two scalars
representing this tensor. The
symmetry of the tensor $\beta^{ls(pq)}$ is similar:
\begin{equation}
\beta^{ls(pq)} = \beta^{sl(pq)} = \beta^{ls(qp)} \,, \label{beta}
\end{equation}
and it also has 36 independent components in general case, and
only two parameters in the spatially isotropic case. Finally, the
tensor $\gamma^{lm(pq)}$ with the symmetry
\begin{equation}
\gamma^{lm(pq)} = \gamma^{lm(qp)} \,, \label{nu}
\end{equation}
is characterized by $9 \times 6 {=} 54$ independent components in
general case, and vanishes in the spatially isotropic medium.
Using (\ref{0nu}), we see that the tensor
$Q^{abmnpq}$
is generally characterized by 126 independent components.

\subsection{Gravity field equations}

The extended gravity field equations
$$
\frac{1}{\kappa} \left[R_{ik}{-}\frac{1}{2}Rg_{ik}\right] =
$$
\begin{equation}
= T^{(0)}_{ik} {+}
T^{({\rm DE})}_{ik} {+} T^{({\rm DM})}_{ik} {+} T^{({\rm m})}_{ik}{+}T^{({\rm W})}_{ik}
{+}T^{({\rm P})}_{ik} {+} T^{({\rm S})}_{ik}
\label{EineqMIN2}
\end{equation}
contain the stress-energy tensor of the electromagnetic field in
the material medium $T^{(0)}_{ik}$ defined by (\ref{TEM}), the
stress-energy tensor of the dark energy  $T^{({\rm DE})}_{ik}$
presented by (\ref{fluid}), the stress-energy tensor of the
standard matter $T^{({\rm m})}_{ik}$ decomposed as
(\ref{Tmatter}), the stress-energy tensor of the pseudoscalar
(axion) field $T^{({\rm DM})}_{ik}$ given by
\begin{equation}
T^{({\rm DM})}_{ik} = \nabla_i \phi \nabla_k \phi - \frac12
g_{ik} \nabla^n \phi \nabla_n \phi + \frac12 g_{ik}
{\cal V}(\phi^2) \,,
\label{axTE}
\end{equation}
and three new interaction terms. The term $T^{({\rm W})}_{ik}$
connected with the pyro-type interactions is of the form
$$
T^{({\rm W})}_{ik} =  DW  \left[\frac12 g_{ik} F_{mn}\left(\pi^{mn}{+} \frac12 F_{ls} \lambda^{mnls}\right)
{-} \right.
$$
$$
\left.
{-}F_{mn}\left(\frac{\delta}{\delta g^{ik}} \pi^{mn} {+} \frac12 F_{ls} \frac{\delta}{\delta g^{ik}} \lambda^{mnls} \right)\right] +
$$
$$
 + \left[W\left(\frac12 g_{ik}+U_iU_k \right) - 2 {\cal B}_{ikls} U^lU^s \right] \times
$$
$$
\times \nabla_j \left[U^j F_{mn}\left(\pi^{mn}{+}
 \frac12 F_{ls} \lambda^{mnls}\right)\right] -
$$
\begin{equation}
- \frac12  F_{mn}\left(\pi^{mn}{+} \frac12 F_{ls}
\lambda^{mnls}\right)U_{(i} \nabla_{k)} W \,. \label{pi195}
\end{equation}
The term $T^{({\rm P})}_{ik}$ relates to the piezo-type contribution to the total stress-energy tensor of the system; it has the form
$$
T^{({\rm P})}_{ik} = \frac12 g_{ik} {\cal D}^{mnpq}F_{mn} {\cal P}_{pq}
{+} 2 {\cal D}^{mnpq}F_{mn} {\cal B}_{ikls}\Delta^l_p \Delta^s_q {-}
$$
\begin{equation}
{-}
F_{mn} {\cal P}_{ls} \frac{\delta}{\delta g^{ik}}\left({\cal D}^{mnpq} \Delta^l_p \Delta^s_q \right)
\,. \label{S99}
\end{equation}
The last new term describes the contribution of the striction -
type interactions; it can be written as
$$
T^{({\rm S})}_{ik} = \frac14 g_{ik} Q^{abmnpq}F_{ab}F_{mn} {\cal P}_{pq} -
$$
$$
-\frac12 F_{ab}F_{mn} {\cal P}_{ls} \frac{\delta}{\delta g^{ik}}\left(Q^{abmnpq} \Delta^l_p \Delta^s_q \right)
+
$$
\begin{equation}
+ Q^{abmnpq}F_{ab} F_{mn} {\cal B}_{ikls}\Delta^l_p \Delta^s_q
\,. \label{S78}
\end{equation}
The tensor ${\cal B}_{ikls}$ in the formulas (\ref{pi195}), (\ref{S99}) and (\ref{S78}) is defined as follows
\begin{equation}
{\cal B}_{ikls} \equiv \frac{1}{\sqrt{-g}} \frac{\delta^2 }{
\delta g^{ik} \delta g^{ls}} \left[\sqrt{-g} \ L_{({\rm
DE})}\right]
 \,. \label{S2}
\end{equation}
This four-indices tensor has the following symmetry:
\begin{equation}
{\cal B}_{ikls}={\cal B}_{kils}={\cal B}_{iksl}={\cal B}_{lsik}\,.
\label{sym1}
\end{equation}
It can be decomposed phenomenologically using the similar
algebraic procedure as for the decomposition of the stress-energy
tensor of the dark energy (\ref{TAX}):
$$
{\cal B}_{ikls}= {\cal G} U_i U_k U_l U_s + \left({\cal G}^{(1)}_{ls} U_i U_k + {\cal G}^{(1)}_{ik} U_l U_s
\right) +
$$
$$
+ \left({\cal G}_i U_k
U_l U_s +{\cal G}_k U_i U_l U_s + {\cal G}_l U_i U_k U_s +{\cal
G}_s U_i U_k U_l \right)+
$$
$$
+  \left({\cal G}^{(2)}_{ks} U_i U_l + {\cal G}^{(2)}_{kl}
U_i U_s + {\cal G}^{(2)}_{is} U_k U_l + {\cal G}^{(2)}_{il} U_k
U_s \right) +
$$
\begin{equation}
+\left({\cal G}_{ikl} U_s + {\cal G}_{iks} U_l + {\cal G}_{kls}
U_i + {\cal G}_{ils} U_k \right) + {\cal G}_{ikls} \,,
\label{sym2}
\end{equation}
where
$$
{\cal G} \equiv  U^aU^b {\cal B}_{abcd} U^cU^d \,, \quad {\cal
G}_s \equiv  U^aU^b {\cal B}_{abcd} U^c \Delta^d_s \,,
$$
$$
{\cal G}^{(1)}_{ls} \equiv  U^aU^b {\cal B}_{abcd} \Delta^c_l \Delta^d_s
\,,
$$
$$
{\cal G}^{(2)}_{ks} \equiv  U^aU^c {\cal B}_{abcd} \Delta^b_k
\Delta^d_s \,, \quad {\cal G}_{kls} \equiv  U^a {\cal B}_{abcd}
\Delta^b_k \Delta^c_l \Delta^d_s \,,
$$
\begin{equation}
{\cal G}_{ikls} \equiv
{\cal B}_{abcd} \Delta^a_i \Delta^b_k \Delta^c_l \Delta^d_s
 \,.
\label{sym3}
\end{equation}
The projected four-indices tensor appeared in (\ref{S99}),
(\ref{S78})
\begin{equation}
{\cal B}_{ikls} \Delta_p^l \Delta_q^s  =  {\cal G}^{(1)}_{pq} U_i
U_k  + {\cal G}_{kpq} U_i + {\cal G}_{ipq} U_k  + {\cal G}_{ikpq}
\label{S221}
\end{equation}
contains only four terms. The two-indices tensor contributed to
(\ref{pi195})
\begin{equation}
{\cal B}_{ikls} U^l U^s  = {\cal G} U_i U_k + {\cal G}_i U_k
 +{\cal G}_k U_i + {\cal G}^{(1)}_{ik}   \,,
 \label{S229}
\end{equation}
also includes only four terms. We will calculate directly the
tensors ${\cal G}$, ${\cal G}_k$, ${\cal G}^{(1)}_{ls}$,
${G}^{(2)}_{ks}$, ${\cal G}_{ikl}$ and ${\cal G}_{ikls}$ below for
the model with spatial isotropy.

\subsection{Equations for the pseudoscalar (axion) field}

Since we represented the dark matter constituent of the dark fluid by a
pseudoscalar (axion) field, we can easily derive the evolutionary equation for
the dark matter by variation of the action functional (\ref{actmin}) with respect
to the pseudoscalar $\phi$; this procedure yields
\begin{equation}
\left[\nabla^k  \nabla_k {+} {\cal V}^{\prime}(\phi^2)\right] \phi
= - \frac{1}{4 \Psi^2_0} F^{*}_{mn}F^{mn} \,.
 \label{axion33}
\end{equation}
In \cite{BBT1,BBT2,BNi} we studied more sophisticated equations describing
the interaction between electromagnetic field and axionic dark matter; nevertheless, here we restrict ourselves by
this simplest model of the axion-photon coupling.

\subsection{Short summary}

We derived the set of coupled master equations for the extended
model: first, electrodynamic equations (\ref{1induc1})-(\ref{C1});
second, gravity field equations (\ref{EineqMIN2})-(\ref{S229});
third, equations for the axion field (\ref{axion33}). Of course,
the derivation of these equations is only the first step in our
program. In the next paper we plan to apply these equations for
the description of anisotropic models of early universe (in
particular, to the Bianchi-I model with global magnetic field) and
to consider an anisotropic dark energy (in analogy with, e.g.,
\cite{DEanis}). For these applications pyro- and piezo- effects
seem to be important. Below we consider only one application of
the established model, namely, the application to the isotropic
homogeneous model of the Friedmann type. We hope it will be a good
illustration that the established model is worthy of attention.

\section{Master equations for a  spatially isotropic medium}

During the late-time universe evolution the dark fluid is
considered as a spatially isotropic substratum. The dark matter is
modelled as a cold substratum with vanishing pressure. The
pressure tensor of such dark fluid is proportional to the
projector, ${\cal P}_{ik} = {-} P \Delta_{ik}$, and the scalar
quantity $P$ describes the pressure of the dark energy. Concerning
the symmetry of tensor coefficients in the context of isotropic
and homogeneous cosmological model, we have to assume that all
pyro- and piezo- coefficients are vanishing. In addition, we have
to assume that all the non-vanishing  tensor coefficients can be
constructed using three basic elements: first, pure geometrical
quantities (metric $g_{ik}$,  Levi-Civita tensor
$\epsilon^{ikmn}$, Kronecker deltas $\delta^i_k$,
$\delta^{ik}_{mn}$, etc.); second, the dynamic quantity $U^i$
(macroscopic velocity of the dark energy), the projector
$\Delta_{ik}$; third, phenomenologically introduced coupling
constants (in front of corresponding terms). We can calculate
directly the variation of all such quantities with respect to the
metric, thus completing the model reconstruction. Let us consider
in detail the model with spatial isotropy.

\subsection{Reduction of the electrodynamic equations}

In the spatially isotropic medium we have to use the tensor of
linear response in the following form:
\begin{equation}
C_{(0)}^{ikmn} {=} \frac{1}{2\mu_{(0)}}\left[g^{ikmn}
{+}(\varepsilon_{(0)}
\mu_{(0)}{-}1)\left(g^{ikmn}{-}\Delta^{ikmn}\right) \right].
\label{C0}
\end{equation}
Here the scalars $\varepsilon_{(0)}$ and $\mu_{(0)}$ are dielectric
and magnetic permittivities of the medium, respectively, in the case when
the dark fluid influence on this medium is negligible. Similarly, the tensor $\lambda^{ikmn}$ is of the form
\begin{equation}
\lambda^{ikmn} {=} \frac{1}{2}\left[\lambda_2 g^{ikmn}
+(\lambda_1 {-} \lambda_2)\left(g^{ikmn}{-}\Delta^{ikmn}\right) \right],
\label{C011}
\end{equation}
where two phenomenological constants $\lambda_1$ and $\lambda_2$ introduce contributions of the
pyro-type interactions (proportional to $DW$) into the
total linear response tensor. Because of spatial isotropy the tensor of pyro-coefficients $\pi^{ik}$
and the tensor of piezo-coefficients ${\cal D}^{ikmn}$ vanish.

The last non-trivial element of the extended theory is the tensor
of striction-type activity $Q^{ikmnpq}$; in order to construct it
we will write, first of all, the space-like tensors
$\alpha^{im(pq)}$ and $\beta^{im(pq)}$ using their symmetry:
\begin{equation}
\alpha^{im(pq)} {=} \alpha_{(1)} \Delta^{im} \Delta^{pq} {+}
\alpha_{(2)} (\Delta^{ip}\Delta^{mq}{+} \Delta^{iq}\Delta^{mp}) \,,
\label{abg1}
\end{equation}
\begin{equation}
\beta^{im(pq)} = \beta_{(1)} \Delta^{im} \Delta^{pq} + \beta_{(2)}
(\Delta^{ip}\Delta^{mq}+ \Delta^{iq}\Delta^{mp}) \,. \label{abg2}
\end{equation}
As for the (pseudo)tensor $\gamma^{im(pq)}$, keeping in mind the
analogy with classical electrodynamics of spatially isotropic
continuous media \cite{Nye,SSh}, we assume that this
(pseudo)tensor of cross-effects is equal to zero, i.e.,
$\gamma^{im(pq)} {=} 0$. In other words, when the medium is
spatially isotropic, we deal with four coupling parameters
$\alpha_{(1)}$, $\alpha_{(2)}$, $\beta_{(1)}$, $\beta_{(2)}$,
which characterize electro- and magneto-striction. The
corresponding reconstruction of the tensor $Q^{ikmnpq}$ yields
$$
Q^{ikmnpq}= \frac12 \left[\alpha_{(1)}\Delta^{pq}
\left(g^{ikmn}{-}\Delta^{ikmn} \right) {+} \right.
$$
$$
\left.{+}
\alpha_{(2)} U_l U_s
\left(g^{iklp} g^{mnsq}{+} g^{iklq}g^{mnsp}\right) + \right.
$$
\begin{equation}
\left. + \beta_{(1)}\Delta^{pq}\Delta^{ikmn} -
\beta_{(2)}(\eta^{ikp}\eta^{mnq}{+}\eta^{ikq}\eta^{mnp})
\right]\,. \label{abg4}
\end{equation}
Clearly, only the four-indices tensor $Q^{ikmnpq} \Delta_{pq}$ enters
the electrodynamic equations, when the pressure tensor of the dark energy is
spatially isotropic; it has now the form
\begin{equation}
Q^{ikmn} \equiv Q^{ikmnpq}\Delta_{pq} = \frac12 \alpha g^{ikmn} + \frac12 (\beta-\alpha)\Delta^{ikmn}
\,, \label{abg49}
\end{equation}
i.e., only two effective coupling constants
\begin{equation}
\alpha = 3 \alpha_{(1)} + 2
\alpha_{(2)} \,,  \quad \beta = 3
\beta_{(1)} + 2 \beta_{(2)} \,, \label{asm22}
\end{equation}
appeared in this tensor instead of four parameters $\alpha_{(1)}$,
$\alpha_{(2)}$, $\beta_{(1)}$ and $\beta_{(2)}$. Total
permittivity tensors of the spatially isotropic medium influenced
by the dark energy are now the following:
\begin{equation}
\varepsilon^{im} = \Delta^{im} \varepsilon \,, \quad \varepsilon =
\varepsilon_{(0)} + \lambda_1 DW - \alpha P\,, \label{e22}
\end{equation}
\begin{equation}
(\mu^{-1})_{ab} = \frac{1}{\mu} \ \Delta_{ab}  \,, \quad
\frac{1}{\mu} = \frac{1}{\mu_{(0)}} + \lambda_2 DW - \beta P \,, \label{m22}
\end{equation}
\begin{equation}
\nu^{am} = 0 \,. \label{n22}
\end{equation}
This means that in the spatially isotropic case one can define the
scalar refraction index of the medium, $n$, accounting for the
influence of the dark energy, yielding
\begin{equation}
n^2 \equiv \varepsilon \mu =  \frac{n^2_{(0)} + \mu_{(0)}(\lambda_1 DW -  \alpha P)}{1 + \mu_{(0)}(\lambda_2 DW - \beta P) } \,, \label{2varco}
\end{equation}
where $n^2_{(0)} \equiv \varepsilon_{(0)} \mu_{(0)}$ is the square
of the refraction index of the medium which does not feel the dark energy influence.
The corresponding phase velocity of the electromagnetic
waves in the striction-active medium is
\begin{equation}
V_{({\rm ph})} \equiv \frac{c}{n}= c \sqrt{\frac{1 + \mu_{(0)}(\lambda_1 DW - \beta P)}{n^2_{(0)} + \mu_{(0)}(\lambda_2 DW - \alpha P)}} \,. \label{v1}
\end{equation}
The group velocity of the electromagnetic waves is defined as
\begin{equation}
V_{({\rm gr})} \equiv c \ \frac{2n}{({n}^2+1)} \,, \label{v2}
\end{equation}
(see, e.g., \cite{BBL} for details), and can be easily displayed
using (\ref{2varco}). Let us mention that the influence of the
dark energy provides the spatially isotropic electrodynamic system
to possess non-stationary properties, when the universe expands.
To be more precise, the refraction index, the phase and group
velocities become functions of the cosmological time: $n(t)$,
$V_{({\rm ph})}(t)$, $V_{({\rm gr})}(t)$, due to the coupling to
the non-stationary dark energy with time dependent energy density
$W(t)$ and pressure $P(t)$.

\subsection{Reduction of the gravity field equations}

In order to reduce the formulas (\ref{pi195})-(\ref{S2}) for the
spatially isotropic case we have to make the following preliminary
steps. First, we put the tensors $\pi^{ik}$, ${\cal D}^{ikpq}$
equal to zero, since now the spontaneous
polarization-magnetization is inadmissible because of the model
symmetry. The second step is to calculate directly the variation
derivatives $ \frac{\delta}{\delta g^{ik}} \lambda^{mnls}$,
$\frac{\delta}{\delta g^{ik}}\left(Q^{abmnpq} \Delta^l_p
\Delta^s_q \right)$ using the reduced formulas (\ref{C011}),
(\ref{abg4}) and the auxiliary formulas, which we presented in
Appendix. The third step is to derive the formulas for the
variation derivatives $ \frac{\delta}{\delta g^{ik}} W$,
$\frac{\delta}{\delta g^{ik}} DW$ and  $\frac{\delta}{\delta
g^{ik}}P$. Let us consider in detail the third step.

Our ansatz is that the {\it spatially isotropic} dark energy can
be modelled by a real scalar field $\Psi$ with the Lagrangian
\begin{equation}
L_{({\rm DE})} = - \frac12 g^{mn} \partial_m \Psi \partial_n \Psi
+ \frac12 V(\Psi^2)\,, \label{DE1}
\end{equation}
and the corresponding stress-energy tensor $T^{({\rm DE})}_{ik}$
\begin{equation}
T^{({\rm DE})}_{ik} {=} \partial_i \Psi \partial_k \Psi {-} \frac12
g_{ik} g^{mn} \partial_m \Psi \partial_n \Psi {+} \frac12 g_{ik}
V(\Psi^2)\,. \label{DE2}
\end{equation}
This idea correlates with the attempt to describe the dark matter
in terms of pseudoscalar field $\phi$ (see (\ref{actmin}) and (\ref{axTE})).
The velocity four-vector $U^i$ is defined as an eigen-vector of
the tensor $T^{({\rm DE})}_{ik}$, and we obtain readily
$$
T^{({\rm DE})}_{ik} U^k = W
U_i=
$$
\begin{equation}
=\partial_i \Psi D \Psi - \frac12 U_i
g^{mn} \partial_m \Psi \partial_n \Psi + \frac12 U_i V(\Psi^2)  \,, \label{DE3}
\end{equation}
\begin{equation}
W \equiv U^iT^{({\rm DF})}_{ik}U^k {=} (D\Psi)^2 {-} \frac12
g^{mn}
\partial_m \Psi \partial_n \Psi {+} \frac12 V(\Psi^2)\,.
\label{DE4}
\end{equation}
Thus, $\partial_n \Psi {=} U_n D\Psi$ and we obtain the well-known
relationships (see, e.g., \cite{Odin}):
$$
W = \frac12 \left[(D\Psi)^2 + V(\Psi^2) \right] \,,
$$
\begin{equation}
P \equiv - \frac13 \Delta^{im} T^{({\rm DF})}_{ik}\Delta^{k}_m  =
\frac12 \left[(D\Psi)^2 - V(\Psi^2) \right]\,, \label{DE5}
\end{equation}
which give (see auxiliary formulas in Appendix)
$$
\frac{\delta}{\delta g^{ik}} W = \frac{\delta}{\delta g^{ik}} P =
\frac12 \partial_i \Psi
\partial_k \Psi =
$$
\begin{equation}
= \frac12 U_i U_k (D\Psi)^2 = \frac12 U_i U_k (W+P) \,.
\label{DE6}
\end{equation}
Similarly, direct calculation of the tensor ${\cal B}_{ikls}$
yields
$$
{\cal B}_{ikls} = \frac14 \left(2 W {-} P \right)U_{i} U_{k} U_{l}
U_{s} +
$$
$$
+ \frac14 W \left(U_{i} U_{k} \Delta_{ls}+ U_{l} U_{s}
\Delta_{ik}\right) - P  U_{(l} \Delta_{s)(k}U_{i)} -
$$
\begin{equation}
 - \frac14 P \left(
\Delta_{ls}\Delta_{ik} {+} \Delta_{li}\Delta_{ks} {+}
\Delta_{lk}\Delta_{is}\right) \,. \label{DE7}
\end{equation}
Thus, for this illustrative example we obtain
$$
{\cal G} =  \frac14 \left(2 W {-} P \right) \,, \quad {\cal
G}_s = 0 \,,
$$
$$
{\cal G}^{(1)}_{ls} = \frac14 W  \Delta_{ls}
\,,
\quad {\cal G}^{(2)}_{ks} = {-} \frac14 P \Delta_{ls} \,, \quad {\cal G}_{kls} = 0  \,,
$$
\begin{equation}
{\cal G}_{ikls} {=}  - \frac14 P \left(
\Delta_{ls}\Delta_{ik} {+} \Delta_{li}\Delta_{ks} {+}
\Delta_{lk}\Delta_{is}\right)
 \,.
\label{2sym3}
\end{equation}
These formulas give the hint to write the working formula
$$
 \frac{\delta}{\delta g^{ik}} {\cal P}_{pq} {=}
$$
\begin{equation}
{=}\frac12 U_iU_k \left[{\cal P}_{pq} {-}\Delta_{pq} W \right] {-} \frac12 \left[g_{p(i} {\cal P}_{k)q} {+} g_{q(i} {\cal P}_{k)p} \right]
, \label{DE8}
\end{equation}
and then to summarize the total stress-energy tensor of the
electromagnetic field interacting with the dark energy as follows:
$$
T^{({\rm EM})}_{ik} = T^{(0)}_{ik} + T^{({\rm W})}_{ik} + T^{({\rm
S})}_{ik} =
$$
$$
= \left[\frac14 g_{ik}F_{mn}-\frac12
\left(g_{im}F_{kn}+g_{km}F_{in} \right) \right]{\cal
C}^{mnab}F_{ab} +
$$
$$
+\frac18 U_i U_k F_{ab}F_{mn}\left\{(W+P)Q^{abmn} + \right.
$$
$$
\left. +
\lambda^{abmn}\left[(W+P)\nabla_j U^j -DW \right]\right\} +
$$
\begin{equation}
 + \frac18 U_i U_k (W+P)
D\left(\lambda^{abmn}F_{ab}F_{mn} \right)
 \,. \label{DE9}
\end{equation}
Here we used the notations
$$
{\cal C}^{abmn} = C_{(0)}^{abmn} +  \lambda^{abmn} \ DW - Q^{abmn}
\ P=
$$
\begin{equation}
= \frac{1}{2\mu}\left[g^{abmn} + \left(\varepsilon \mu -1\right)
\left(g^{abmn} - \Delta^{abmn}\right)\right] \,, \label{DE10}
\end{equation}
where the quantities $\mu$  and $\varepsilon$ are already defined by
(\ref{e22}) and (\ref{m22}).

\subsection{Model with hidden anisotropy}

The spatial isotropy of the space-time happens to be  violated,
when one considers the model with global magnetic and/or electric fields.
For the magnetic field in vacuum one should study the Bianchi-I
cosmological model instead of the Friedmann one, since the tensor
$\Delta^p_iT_{pq}^{({\rm EM})}\Delta^q_k$ in the right-hand side
of the Einstein equations is not spatially isotropic in this case
(i.e., $\Delta^{1p}T_{pq}^{({\rm EM})}\Delta^q_1 {=}\Delta^{2p}T_{pq}^{({\rm EM})}\Delta^q_2 \neq \Delta^{3p}T_{pq}^{({\rm EM})}\Delta^q_3$,
when $x^3$ is the anisotropy axis).
When an electrodynamic system interacts with dark energy, the
situation changes essentially: one can find such states of the
system, for which the magnetic field is non-vanishing but the
spatial isotropy is inherited. For instance, when
\begin{equation}
\frac{1}{\mu_{(0)}} + \lambda_2 DW -\beta P =0 \,, \label{TEM121}
\end{equation}
we obtain immediately that
$$
\Delta^p_iT_{pq}^{({\rm EM})}\Delta^q_k =
$$
\begin{equation}
= (\varepsilon_{(0)}{+}\lambda_1 DW{-}\alpha
P)\left[\frac{1}{2}\Delta_{ik} E_m E^m {-} E_i E_k  \right] \,,
\label{TEM112}
\end{equation}
where $E^i{=}F^{ik}U_k$ is the electric field four-vector; for
pure magnetic field $E^i{=}0$, thus $\Delta^p_iT_{pq}^{({\rm
EM})}\Delta^q_k {=}0$. This model with {\it hidden} magnetic
anisotropy is exotic, since the effective refraction index is now
equal to infinity, and the phase and group velocities of the
electromagnetic waves are equal to zero for such dark medium.
Similarly, when $\varepsilon_{(0)} {+} \lambda_1 DW {-}\alpha P
{=}0$, but $\frac{1}{\mu_{(0)}}\neq \beta P {-}\lambda_2 DW $ we
deal with the so-called hidden electric anisotropy. Finally, when
$\varepsilon_{(0)}{=}\alpha P{-}\lambda_1 DW$ and
$\frac{1}{\mu_{(0)}} {=} \beta P{-}\lambda_2 DW$ simultaneously,
the stress-energy tensor $\Delta^p_iT_{pq}^{({\rm EM})}\Delta^q_k
$ is equal to zero identically, and the electromagnetic source of
the gravity field disappears from the Einstein equations. Similar
results related to a hidden magnetic anisotropy were obtained
early in the frameworks of the nonminimal Einstein-Maxwell theory
\cite{BaZim} and extended Einstein-Maxwell theory
\cite{2007A,2007B}.

\section{Cosmological applications: Unlighted epochs}

We obtained extended master equations for the coupled
electromagnetic and gravitational fields, which take into account
interactions of pyro-, piezo- and striction- types. In the nearest
future we intend to analyze applications of these master equations
to the early universe with global magnetic field, and to the
problem of late-time universe accelerated expansion. In this work
we consider only one illustration of the extended model. To be
more precise, in this Section we assume that the space-time is of
the Friedmann-Lema\^itre-Robertson-Walker type with the metric
\begin{equation}
ds^2 = c^2dt^2 - a^2(t)(dx^2+dy^2+dz^2) \,, \label{F1}
\end{equation}
and this space-time is a fixed background for a local (test)
electrodynamic system. In other words, here we neglect by the
backreaction of the electromagnetic field on the gravity field,
but consider the striction-type influence of the spatially
isotropic cosmic dark energy on the electrodynamic system. We are
interested in the analysis of the so-called unlighted epochs in
the universe history, analogs of which were described in
\cite{BBL} in the framework of nonminimal field theory. We use the
term "unlighted epochs" for intervals of the universe evolution,
for which the square of the effective refraction index $n^2(t)$
(see (\ref{2varco})) takes negative values, $n^2(t)<0$. During
these periods of time the refraction index is a pure imaginary
quantity, and the phase and group velocities of the
electromagnetic waves (see (\ref{v1})) and (\ref{v2}) are not
defined. Clearly, the function $n^2(t)$ can change sign at the
moments $t_{(s)}$ of the cosmological time when the numerator or
the denominator in (\ref{2varco}) vanish. When the numerator
vanishes, one has $n(t_{(s)}){=}0$, $V_{({\rm
ph})}(t_{(s)}){=}\infty$, and $V_{({\rm gr})}(t_{(s)}){=} 0$. When
the denominator vanishes, one has $n(t_{(s)}){=}\infty$, $V_{({\rm
ph})}(t_{(s)}){=}0$, and $V_{({\rm gr})}(t_{(s)}){=} 0$. In both
cases the unlighted epochs appear and disappear when the group
velocity of the electromagnetic waves vanishes, i.e., at these
points the energy transfer stops. We indicate the times $t_{(s)}$
as the unlighted epochs boundary points (see \cite{BBL} for
details). Below we consider three very illustrative examples of
the evolution of the cosmic dark energy and of the corresponding
behavior of the pressure function $P(t)$ for the case, when the
striction coefficients only are non-vanishing.

\subsection{De Sitter-type models}

The simplest model of the dark energy is the de Sitter one;
for this model the dark energy pressure is constant $P{=}{-} \Lambda$.
Clearly, the refraction index for this model is also constant
\begin{equation}
n^2 = \frac{n^2_{(0)} + \mu_{(0)} \alpha \Lambda}{1 + \mu_{(0)}
\beta \Lambda } \,, \label{ds1}
\end{equation}
i.e., unlighted epochs are not available. When
\begin{equation}
\beta - \alpha = \frac{n^2_{(0)}-1}{\mu_{(0)}  \Lambda} \,,
\label{ds18}
\end{equation}
we obtain that the universe expansion is characterized by the
condition $n^2{=}1$, so that $V_{({\rm gr})} {=} V_{({\rm ph})}
{=} c$.

\subsection{Anti-Gaussian solution}

 In \cite{1BB2011,2BB2011} the exact solution of the
Archimedean-type model is obtained, which was indicated as
Anti-Gaussian solution, since for this solution the scale factor
$a(t)$ is of the form
\begin{equation}
a(t) = a(t^*) \exp{\left[\frac{8\pi G}{3\nu}(t-t^*)^2\right]}\,.
\label{pi2}
\end{equation}
Here we repeated the notations from \cite{1BB2011}
$$
\log{\frac{a(t^*)}{a(t_0)}} \equiv - \frac{\nu}{4}[\rho(t_0)+
E_{(0)}] \,,
$$
\begin{equation}
t^* = t_0 - \nu \sqrt{\frac{3}{32\pi
G}[\rho(t_0)+ E_{(0)}]} \,. \label{pi4}
\end{equation}
The parameter $\nu$ is a coupling constant of the Archimedean-type
interaction between dark energy and dark matter, the parameters
$a(t_0)$, $\rho(t_0)$ and $E_{(0)}$ are the initial data for the
scale factor, energy-density of the dark energy and energy-density
of the dark matter, respectively. According to that model the dark
energy pressure and energy-density (here and below we use the symbols $\Pi(t)$ and $\rho(t)$, respectively, for these
quantities in analogy with \cite{1BB2011,2BB2011}) are described by the formulas
$$
\Pi(t)= \Pi(t_0) - \frac{4}{\nu} \log{\left[\frac{a(t)}{a(t_0)}\right]} \,,
$$
\begin{equation}
\rho(t)= \rho(t_0) + \frac{4}{\nu} \log{\left[\frac{a(t)}{a(t_0)}\right]} \,.
\label{pi1}
\end{equation}
The formula for the pressure can be rewritten in the form
$$
\Pi(t) = \Pi(t^*) - \frac{32\pi G}{3\nu^2}(t-t^*)^2\,,
$$
\begin{equation}
\Pi(t^*) = \Pi(t_0) + \rho(t_0) + E_{(0)} \,. \label{pi3}
\end{equation}
We assume that the late-time universe evolution is characterized
by the model with $n^2(t) \to 1$; this assumption provides that at
present the light propagates with phase and group velocities equal
to the speed of light in vacuum. According to (\ref{2varco}) and
(\ref{pi3}) this requirement at $t \to \infty$ leads to the
equality $\beta {=} \alpha$. Also, we put $\mu_{(0)}{=1}$ for
simplicity and assume that $n^2_{(0)}{=}\varepsilon_{(0)}>1$. Then
we use the replacement
\begin{equation}
z^2 \equiv \frac{32\pi G}{3\nu^2}(t-t^*)^2
 \label{pi33}
\end{equation}
and transform (\ref{2varco}) into
\begin{equation}
n^2 = \frac{z^2-Z_2}{z^2-Z_1} \,, \label{pi5}
\end{equation}
where
$$
Z_2 \equiv \Pi(t^*) - \frac{n^2_{(0)}}{\alpha} \,, \quad Z_1
\equiv \Pi(t^*) - \frac{1}{\alpha} \,,
$$
\begin{equation}
\alpha \left(Z_1-Z_2
\right) = [n^2_{(0)}-1] >0 \,. \label{pi6}
\end{equation}
Now we are ready to describe unlighted epochs.

\subsubsection{Models without unlighted epochs}

The refraction index $n(t)$ is equal to one identically, when
$n^2_{(0)}{=}1$ and thus $Z_1{=}Z_2$. Also, the quantity $n^2(t)$ is
positive for {\it arbitrary} time, when $Z_1$ and $Z_2$ are
negative, i.e.,
\begin{equation}
\Pi(t^*) < \frac{n^2_{(0)}}{\alpha} \,, \quad \Pi(t^*) <
\frac{1}{\alpha} \,. \label{pi689}
\end{equation}
In both cases the unlighted epochs can not appear.

\subsubsection{Unlighted epochs of the first type}

Let the parameter $Z_1$ be negative, and $Z_2$ be positive. Clearly, it is possible, when
\begin{equation}
\frac{n^2_{(0)}}{\alpha} < \Pi(t^*) < \frac{1}{\alpha} \ \ \Rightarrow \ \ \alpha <0  \,, \quad \Pi(t^*) <0 \,.
\label{un1}
\end{equation}
Thus, the refraction index is imaginary, when $|z|< \sqrt{Z_2}$, or in more details,
\begin{equation}
|t-t^*| < \Delta_2 \,, \quad \Delta_2 \equiv \sqrt{\frac{3\nu^2}{32\pi G}\left[\Pi(t^*)- \frac{n^2_{(0)}}{\alpha} \right]}\,.
\label{un2}
\end{equation}
At the boundary points of this time interval, $t_{(\pm)}{=}t^* \pm \Delta_2$, the refraction index and the group velocity
vanish, $n(t_{(\pm)}){=}0$, $V_{({\rm gr})}(t_{(\pm)}){=}0$, and the phase velocity becomes infinite $V_{({\rm ph})}(t_{(\pm)}){=} \infty$.
The duration of this unlighted epoch is equal to $2\Delta_2$ (see Panel A of Fig.1).

\begin{figure}[h]
\centerline{\includegraphics[width=8cm]{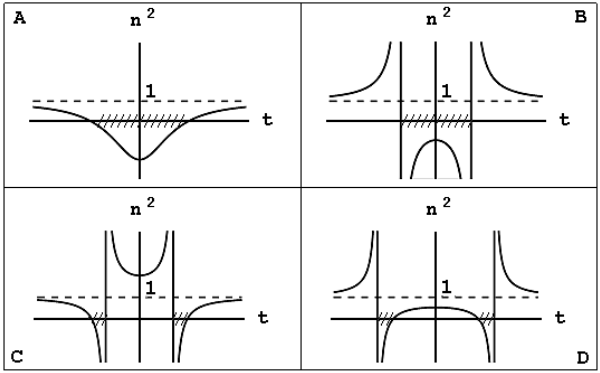}}
\caption{{\small Sketches of basic graphs illustrating unlighted epochs of four types.
Unlighted epochs relate to the intervals of cosmological time  $t$ for which the function $n^2(t)$, the squared effective refraction index, is negative.
Panel A relates to the case, when $Z_1<0$, $Z_2>0$, and thus the denominator of the function $n^2(t)$ is positive (see (\ref{pi33})-(\ref{pi6}));
this panel illustrates the simply connected unlighted epoch of the first type with zeroth values of the function $n^2(t)$ at the boundary points. Panel B illustrates the simply connected unlighted epoch of the second type, which corresponds to the case $Z_1>0$, $Z_2<0$, so that both boundary values of the function $n^2(t)$ are infinite. The graphs of the unlighted epochs of the third type (Panel C) and of the fourth type (Panel D) are doubly-connected; they correspond to the cases $Z_2>Z_1>0$ and $Z_1>Z_2>0$, respectively. When $n^2{=}0$ or $n^2 {=}\infty$, the group velocity of electromagnetic wave (see (\ref{v2})) takes zero value, i.e., the energy transfer stops.  For all the cases one has that $n^2(t\to \pm \infty) \to 1$ (see (dashed) horizontal asymptotes.)}}
\end{figure}

\subsubsection{Unlighted epochs of the second type}

Now, let the parameter $Z_2$ be negative, and $Z_1$ be positive. Clearly, it is possible, when
\begin{equation}
\frac{1}{\alpha} < \Pi(t^*) < \frac{n^2_{(0)}}{\alpha} \ \ \Rightarrow \ \ \alpha >0  \,, \quad \Pi(t^*) > 0 \,.
\label{un3}
\end{equation}
Thus, the refraction index is imaginary, when $|z|< \sqrt{Z_1}$, or in more details,
\begin{equation}
|t-t^*| < \Delta_1 \,, \quad \Delta_1 \equiv \sqrt{\frac{3\nu^2}{32\pi G}\left[\Pi(t^*)- \frac{1}{\alpha} \right]}\,.
\label{un4}
\end{equation}
At the boundary points of this time interval, $t_{(\pm)}{=}t^* \pm \Delta_1$, the refraction index is infinite,
$n(t_{(\pm)}){=} \infty$, thus the group and phase velocities vanish, $V_{({\rm gr})}(t_{(\pm)}){=}0$, $V_{({\rm ph})}(t_{(\pm)}){=} 0$.
The duration of this unlighted epoch is equal to $2\Delta_1$ (see Panel B of Fig.1).

\subsubsection{Unlighted epochs of the third type}

Now we consider the case, when both parameters $Z_1$ and $Z_2$ are positive. For positive $\alpha$ this gives the conditions
\begin{equation}
\Pi(t^*) > \frac{n^2_{(0)}}{\alpha}  \,,
\label{un5}
\end{equation}
so that $n^2(t)<0$, when $\sqrt{Z_2} <|z|< \sqrt{Z_1}$, or equivalently,
\begin{equation}
\Delta_2 < |t-t^*| < \Delta_1  \,.
\label{un6}
\end{equation}
This unlighted epoch is divided into two separated sub-epochs, the duration of both sub-epochs is $\Delta_1{-}\Delta_2$ (see Panel C of Fig.1).
Similarly, when $\alpha<0$ and $\Pi(t^*) > \frac{1}{\alpha}$, we could find two unlighted sub-epochs at $\Delta_1 < |t-t^*| < \Delta_2$ (see Panel D of Fig.1).

\subsection{Super-exponential solution}

In \cite{1BB2011} the exact solution of the
Archimedean-type model is obtained, which was indicated as
super-exponential, since for this solution the scale factor
is of the form
\begin{equation}
\frac{a(t)}{a(t_0)} {=} \exp \left\{ \sqrt{\frac{2 \rho (t_0)}{9 \rho_0}} \sinh{\left[ \sqrt{12 \pi G \rho_0} (t{-}t_0)\right]} \right\},
\label{se1}
\end{equation}
where the parameter $\rho_0$ is the so-called bag constant. The corresponding dark energy pressure is
$$
\Pi(t)= \Pi(t_0) +3\left[\rho(t_0) + \Pi(t_0) - \rho_0 \right] \log{\left[\frac{a(t)}{a(t_0)}\right]} -
$$
\begin{equation}
-
\frac92 \rho_0 \log^2{\left[\frac{a(t)}{a(t_0)}\right]} \,.
\label{se2}
\end{equation}
Surprisingly, this model can be reduced to the one considered in the previous section, if we use the following definition for $z$:
$$
z(t) \equiv  \sqrt{\rho(t_0)} \sinh{\left[ \sqrt{12 \pi G \rho_0} \ (t-t_0)\right]} -
$$
\begin{equation}
-
\frac{\left[\rho(t_0) +
\Pi(t_0) - \rho_0 \right]}{\sqrt{2\rho_0}} \,.
\label{se4}
\end{equation}
In order to complete the analogy, we find the parameter $t^*$ from the equation $z(t^*) {=} 0$ (see (\ref{se4}))
and obtain from (\ref{se2}) and (\ref{se1}) that
\begin{equation}
\Pi(t^*)\equiv \Pi(t_0) + \frac{\left[\rho(t_0) + \Pi(t_0) - \rho_0 \right]^2}{2\rho_0} \,,
\label{se5}
\end{equation}
with $Z_1$ and $Z_2$ inherited from (\ref{pi6}). Thus, we deal again with  the analysis of the formula (\ref{pi5})
and obtain the similar results, if we make the replacement
$$
\sqrt{\frac{32\pi G}{3\nu^2}}(t-t^*) \ \ \Rightarrow \ \
$$
\begin{equation}
\left\{\sqrt{\rho(t_0)} \sinh{\left[ \sqrt{12 \pi G \rho_0}
(t{-}t_0)\right]} {-} \frac{\left[\rho(t_0) {+} \Pi(t_0) {-}
\rho_0 \right]}{\sqrt{2\rho_0}} \right\}. \label{se6}
\end{equation}
Again, the model admits the existence of unlighted epochs of four
types sketched on Panels 1-4 of Fig.1.

\subsection{Remarks on stability of the dark energy scalar potential under quantum fluctuations, and constraints on the striction model}

We established pure classical model of striction-type interactions between dark energy and electrodynamic system.
The paper context does not allow us to consider quantum aspects of this model, however, we hope to return to this problem in the
next work. Here we would like to discuss briefly only three remarks, which could be important for physical understanding of the model consequences.

\subsubsection{How the corrections to the scalar field potential can influence the master equations of the striction-type model? }

The crucial point of establishing the total set of master equations of the striction-type model is the finding of the tensor ${\cal B}_{ikls}$ (see (\ref{S2}) and (\ref{sym2})). Generally, it can not be presented in an explicit form by means of the energy density $W$ and of the dark energy pressure $P$, and its reconstruction requires sophisticated phenomenological decomposition (\ref{sym2}). However, when we treat the dark energy as a spatially homogeneous isotropic medium using an analogy with some scalar field $\Psi$, we obtain ${\cal B}_{ikls}$ by direct variation procedure (see (\ref{DE7})), thus providing the model to be self-closed. Moreover, we find that, when we restrict our-selves by the striction-type interactions only, i.e., $\lambda^{ikmn}{=}0$, the electrodynamic equations include the scalar $P$ only, and the gravity field equations include $P$ and the combination $W{+}P$ only. It follows from (\ref{DE5}) that the sum $W{+}P{=}{(D\Psi)}^2$ does not feel the shape of the potential $V(\Psi^2)$; as for the dark energy pressure $P$, it, clearly, depends on $V(\Psi^2)$. This fact confirms that modeling of striction-type extensions of master equations for the electromagnetic and gravitational fields by means of a scalar dark energy is sensitive to the choice of the scalar field potential.

\subsubsection{One-loop corrections to the scalar field potential \\ in the framework of anti-Gaussian model}

Based on the method of one-loop corrections to potentials of complex or real scalar fields attributed to the dark energy,
dark matter or dark fluid, the authors of the works \cite{loop1,loop2} have shown that these scalar field potentials are
usually stable under quantum fluctuations, however, a coupling to fermions is very restricted. When we consider an electrodynamic
system influenced by a scalar dark energy, we have to take into account these results keeping in mind two aspects.
First, the scalar field potential $V(\Psi^2)$ attributed to the dark energy (see (\ref{DE1})) can be modified due to one-loop corrections along the line discussed in \cite{loop1,loop2};
since this potential enters the formula for the pressure of the dark energy (see (\ref{DE5})), these corrections can re-define the coupling constants of the
striction-type interaction discussed above. Second, the electrodynamic system inevitably contains electrically charged fermions; when the fermion mass
$m_{\rm f}$ depends on the scalar field, $m_{\rm f}(\Psi)$, we can use directly the method and estimations discussed in \cite{loop1,loop2}. As for (possible) dependence of the photon mass on the scalar field $\Psi$, this problem is worthy a special attention and will be discussed in a separate paper.

Let us apply the method used in \cite{loop1,loop2} to the anti-Gaussian model discussed in Subsection V.B. According to (\ref{pi1}) and  (\ref{DE5})), we readily obtain that
\begin{equation}
{\dot{\Psi}}^2(t) = \rho(t) {+} \Pi(t)= \rho(t_0) {+} \Pi(t_0)  \equiv {\dot{\Psi}}^2(t_0) \,,
\label{loop1}
\end{equation}
thus providing the scalar field $\Psi$ to be linear function of the cosmological time $t$:
\begin{equation}
\Psi(t) = \Psi(t_0) + {\dot{\Psi}}(t_0) \ t \,.
\label{loop2}
\end{equation}
Then, using (\ref{DE4})) and (\ref{DE5})) we see that the potential $V(\Psi^2){=} V^{*}(\Psi)$ is a quadratic function of the scalar field:
\begin{equation}
V^*(\Psi) = A + 2B \Psi + C \Psi^2 \,,
\label{loop3}
\end{equation}
where
$$
A = \rho(t_0)-\Pi(t_0) - 2B \Psi(t_0) - C  \Psi^2(t_0) \,,
$$
$$
B = \frac{16}{\nu {\dot{\Psi}}(t_0)} \sqrt{\frac{\pi G}{6}\left[\rho(t_0){+}E_{(0)} \right]} - C \Psi(t_0) \,,
$$
\begin{equation}
C = \frac{64 \pi G}{3 \nu^2 {\dot{\Psi}}^2(t_0)} \,.
\label{loop4}
\end{equation}
For the function (\ref{loop3}) the second derivative of the potential is $V^{*\prime \prime}(\Psi){=} 2C$, thus the formula (2) from  \cite{loop1} gives us the one-loop modification of the scalar potential (\ref{loop3})
\begin{equation}
V_{\rm 1-loop}(\Psi) =  V^{*}(\Psi) {+}  \frac{4 G \Lambda^2_{\rm s}}{3 \pi \nu^2 {\dot{\Psi}}^2(t_0)} {-} \frac{\Lambda^2_{\rm f}}{8\pi^2}[m_{\rm f}(\Psi)]^2
\,,
\label{loop5}
\end{equation}
where $\Lambda_{\rm s}$ and $\Lambda_{\rm f}$ are the ultra-violet cutoffs of the scalar and fermion fluctuations, respectively; $m_{\rm f}(\Psi)$ is $\Psi$-dependent fermion mass, which appears in the Lagrangian of matter $L_{({\rm m})}$ (see (\ref{0actmin})), when the matter is considered to consist of fermions (see (1) in \cite{loop1}).  Clearly, being constant, the second term in the right-hand side of (\ref{loop5}) can be absorbed by the constant $A$ appeared in the potential (\ref{loop3}); the authors of the works \cite{loop1,loop2} indicate such a case as describing the potential stable under scalar fluctuations. As for estimations of the potential stability related to fermion fluctuations, the third term in (\ref{loop5}) is exactly the same as the one in (2) of the paper \cite{loop1}; this means that estimations discussed in Section III of \cite{loop1} are also valid. Let us repeat, that estimations related to the scalar field coupling to photons and gravitons based on one-loop corrections require a special consideration and, unfortunately, are out of frame of this paper.

\subsubsection{How the corrections to the scalar field potential could be visualized in the cosmic striction-type phenomena? }

Let us consider the consequences of the scalar field potential modification, given  by the second term in the right-hand side of (\ref{loop5}), for the unlighted epochs formation.
In fact, this constant one-loop correction to the quadratic potential leads to the following formal re-definition of the parameter $\Pi(t^{*})$ in (\ref{pi3}):
\begin{equation}
\Pi(t^{*}) \to \tilde{\Pi}(t^{*}) = \Pi(t^{*}) - \frac{2 G \Lambda^2_{\rm s}}{3 \pi \nu^2 {\dot{\Psi}}^2(t_0)} \,.
\label{loop6}
\end{equation}
Then we have to replace the parameter $\Pi(t^{*})$ with this new parameter $\tilde{\Pi}(t^{*})$ in all inequalities, which predetermine the structure of the unlighted epochs
(see Sections V.B1, V.B2, V.B3, V.B4). Let us illustrate the consequences of such pressure re-definition by the examples, for which $\alpha >0$ and $\Pi(t^{*})$ is positive
(see, e.g., (\ref{pi689}), (\ref{un3}), (\ref{un5})).
Since the second term in (\ref{loop6}) is negative, we can imagine that $\tilde{\Pi}(t^{*})$ becomes negative due to such scalar potential correction. This means that the inequalities (\ref{pi689}) (with the replacement $\Pi(t^{*}) \to \tilde{\Pi}(t^{*})$) remain valid, so that this veto for the unlighted epoch appearance holds out. The inequalities (\ref{un3}) and  (\ref{un5})) are not valid now, therefore, the corresponding unlighted epochs of the second and third types become forbidden because of scalar fluctuations. In other words, the answer on the question about absence or presence of the unlighted epochs seems to be very sensitive to the choice of the scalar field potential and to its one-loop corrections.

\section{Discussion and conclusions}

1. A new model of coupling between a cosmic dark fluid and
electrodynamic systems is established, i.e., the extended
Lagrangian is introduced and the extended master equations are
derived for electromagnetic and gravitational fields. What is the
novelty of this model from the mathematical point of view? First,
we introduced four cross-terms into the Lagrangian, which contain
the Maxwell tensor up to the second order, on the one hand, and
contain the pressure tensor of the dark energy and the convective
derivative of its energy-density scalar, on the other hand. Thus,
a modified Lagrangian is not of pure field-type, since these
cross-terms are the products of pure field-type elements
($F_{ik}$) and of quantities defined algebraically (see
(\ref{fluid}), (\ref{1fluid}) for the definitions of $U^i$, $W$,
${\cal P}_{pq}$). Second, we described a modified procedure of
variation with respect to the metric, which happened to be
necessary for these new cross-terms defined algebraically. This
modified procedure is based on the rule of variation of the
macroscopic velocity four-vector (\ref{11T1}) taken from the works
\cite{2007A,2007B}; on the rule of algebraic decomposition of the
second variation of the dark energy Lagrangian
(\ref{S2})-(\ref{sym3}), and on the ansatz about the structure of
this decomposition in the case of spatially isotropic medium (see
(\ref{DE6})-(\ref{DE8})).

\vspace{1mm} \noindent 2. What is the physical motivation of the
Lagrangian extension, which we made? Since we follow the
mathematical scheme, which is well-known in the relativistic
electrodynamics of continuous media, we indicate the new
cross-terms in the extended Lagrangian (\ref{actmin}) using the
similar terminology: as (dark energy inspired) analogs of terms
describing electric and magnetic striction, piezo-electricity and
piezo-magnetism, pyro-electricity and pyro-magnetism. This analogy
allowed us to interpret new coupling constants as (dark energy
induced) pyro-, piezo- and striction coefficients, respectively.

\vspace{1mm} \noindent 3. First cosmological application of the
model shows that a striction-type interaction between the dark
energy and test electrodynamic system provides the phase and group
velocities of electromagnetic waves to become sophisticated
functions of cosmological time. In the asymptotic regime, at $t
\to \infty$, these functions tend to the speed of light in vacuum,
i.e., $V_{({\rm ph})} \to c$ and $V_{({\rm gr})} \to c$. However,
during the universe evolution the so-called unlighted epochs can
appear, for which the effective refraction index of the cosmic
medium is an imaginary quantity. At the boundary points of these unlighted epochs
the group velocity of the electromagnetic waves takes zero value, so the electromagnetic energy
transfer stops.

\vspace{1mm} \noindent 4. The last unlighted epoch (if such epochs have ever existed in the real Universe)
should finish before the so-called recombination era, as far as, cosmic microwave background composed of relic photons traveling freely is observable.
In our terms this means that the characteristic time $t_{(\rm last UE)}{=}t^{*}{+}\Delta_2$  is less than $t_{({\rm rec})} \simeq  10^{13} {\rm sec}$.
Using (\ref{pi4}), (\ref{un2}), (\ref{pi3}) at $t_0{=}t_{({\rm rec})}$ and the condition $t_{(\rm last UE)}< t_{({\rm rec})}$, we obtain a cosmological
constraint $\Pi(t_{({\rm rec})})<\frac{n^2(t_{({\rm rec})})}{\alpha}$ linking the phenomenological parameter of a striction-type coupling $\alpha$, the
value of the refraction index $n(t_{({\rm rec})})$ at the end of the recombination era, and the value of the dark energy pressure $\Pi(t_{({\rm rec})})$ at that moment.
It is only one example of constraints appearing in this model; we intend to discuss other constraints in a special note.

\vspace{1mm} \noindent 5.
The criteria of existence of the unlighted epochs are very sensitive to the choice of the scalar field potential, which one uses for modeling of the dark energy.
We have illustrated this sentence on the example of the anti-Gaussian model. In particular, taking into account one-loop corrections to the dark energy
scalar field potential, one can show, that scalar fluctuations are able to avoid the unlighted epoch formation.

\vspace{1mm} \noindent 6.
We expect that the established model can
provide new interesting results in application to the anisotropic
Bianchi-I cosmological model, since in this period of the universe
evolution pyro-magnetic  and piezo-magnetic effects can appear in
addition to the magneto- striction effect admissible both on the
anisotropic and isotropic stages of the universe expansion.

\vspace{3mm}
\noindent
{\bf Acknowledgments}

\noindent
This work is supported by the Russian Foundation for
Basic Research (Grant No. 14-02-00598).

\section*{Appendix}

In order to calculate directly the variation derivatives of basic
quantities, we have to fix the following auxiliary formulas. We
start with the well-known formulas for the variation of the
determinant of the metric and metric itself:
$$
\frac{1}{\sqrt{-g}}\frac{\delta \sqrt{-g}}{\delta g^{ik}} =
-\frac12 g_{ik} \,,
$$
\begin{equation}
\frac{\delta g_{pq}}{\delta g^{ik}} = -
\frac12 \left[g_{p(i} g_{k)q} + g_{q(i} g_{k)p}\right] \,,
 \label{01T1}
\end{equation}
and for the four-indices tensor $g^{ikmn} \equiv
g^{im}g^{kn}{-}g^{in}g^{km}$
\begin{equation}
\frac{\delta g^{abmn}}{\delta g^{ik}} = \delta^{[a}_{(i} g_{k)}^{\
b]mn} +  g_{\ \ \ \ (k}^{ab[m} \delta^{n]}_{i)} \,. \label{1T6}
\end{equation}
The formulas for the variation of the velocity four-vector
$$
 \frac{\delta U^l}{\delta g^{ik}} =  \frac14 \left(\delta^{l}_i U_k {+} \delta^{l}_k U_i \right) = \frac12 \delta^l_{(i} U_{k)} \,,
$$
\begin{equation}
\frac{\delta U_a}{\delta g^{ik}} = - \frac14
\left(g_{ia}U_k {+} g_{ka}U_i \right) = - \frac12 g_{a(i}U_{k)}
\,,
 \label{11T1}
\end{equation}
stand to keep the normalization condition $g^{ik}U_i U_k {=}1$
(see, e.g., \cite{2007A,2007B}). Using (\ref{01T1}) and
(\ref{11T1}) it is easy to write the formulas for the variation of
the projectors
$$
 \frac{\delta \Delta^{pq}}{\delta g^{ik}} = \delta^{(p}_{(i} \Delta^{q)}_{k)} \,,
$$
\begin{equation}
\frac{\delta \Delta_{pq}}{\delta g^{ik}} = - \frac12 \left[g_{p(i} \Delta_{k)q} + g_{q(i} \Delta_{k)p} \right] \,,
 \label{1T2}
\end{equation}
\begin{equation}
 \frac{\delta \Delta^{p}_{q}}{\delta g^{ik}} = \frac12 \left[\delta^p_{(i} \Delta_{k)q} - \Delta^p_{(i} g_{k)q}\right] \,.
 \label{1T3}
\end{equation}
For the variation of the four-indices projector $\Delta^{ikmn}
\equiv \Delta^{im}\Delta^{kn}{-} \Delta^{in}\Delta^{km}$ we  use
the convenient formula
\begin{equation}
 \frac{\delta \Delta^{abmn}}{\delta g^{ik}} =  \delta^{[a}_{(i} \Delta_{k)}^{\ b]mn} +  \Delta_{\ \ \ \ (k}^{ab[m} \delta^{n]}_{i)} \,.
 \label{1T7}
\end{equation}
Finally, the variation of the Levi-Civita tensor $\epsilon^{abpq}$
and of the associated tensor $\eta^{abp} \equiv \epsilon^{abpq}U_q
$ yield
$$
 \frac{\delta \epsilon^{abpq}}{\delta g^{ik}} = \frac12 \epsilon^{abpq} g_{ik} \,,
 $$
 \begin{equation}
  \frac{\delta \eta^{abp}}{\delta g^{ik}} = \frac12 \left[\eta^{abp} g_{ik} - \epsilon^{abp}_{\ \ \ (i}U_{k)} \right] \,.
 \label{1T40}
\end{equation}

\end{document}